\documentclass[12pt,psamsfonts]{amsart}
\usepackage{amsmath}
\usepackage{amsthm}
\usepackage{amssymb}
\usepackage{amscd}
\usepackage{amsfonts}
\usepackage{amsbsy}
\usepackage{epsfig}

\newcommand{\Z}{\ensuremath{\mathbb{Z}}}

\begin{document}

\title[On the origin of the Quantum mechanics]
{On the origin of the Quantum mechanics}

\author[J. Gin\'{e}]
{Jaume Gin\'e}

\address{Departament de Matem\`atica, Universitat de Lleida,
Av. Jaume II, 69. 25001 Lleida, Spain}

\email{gine@eps.udl.es}

\thanks{The author is partially supported by a DGICYT
grant number BFM 2002-04236-C02-01 and by DURSI of Government of
Catalonia ``Distinci\'o de la Generalitat de Catalunya per a la
promoci\'o de la recerca universit\`aria".}


\date{}
\dedicatory{}

\maketitle

\begin{abstract}
Action at distance in Newtonian physics is replaced by finite
propagation speeds in classical post--Newtonian physics. As a
result, the differential equations of motion in Newtonian physics
are replaced by functional differential equations, where the delay
associated with the finite propagation speed is taken into
account. Newtonian equations of motion, with post--Newtonian
corrections, are often used to approximate the functional
differential equations. Are the finite propagation speeds the
origin of the quantum mechanics? In this work a simple atomic
model based on a functional differential equation which reproduces
the quantized Bohr atomic model is presented. As straightforward
application of the result the fine structure of the hydrogen atom
is approached.
\end{abstract}

\section{Introduction}\label{s1}

Newtonian forces (for example, the inverse square law for
gravitation) imply ``action at distance". This absurd, but
outstandingly successful, premise of Newtonian theory predicts
that signals propagate instantaneously. In classical physics,
relativity theory postulates that signals propagate with a
velocity that does not exceed the velocity of light. Thus, the
forces of Newtonian physics must be replaced by force laws that
take into account the finite propagation speed of the classical
fields which determine the forces acting on a moving body. In
turn, the ordinary or partial differential equations of Newtonian
physics, which are derived from the second law of motion $m
\ddot{r} = F$, must be replaced by corresponding functional
differential equations where the force $F$ is no longer a function
of just position, time, and velocity; rather, the classical force
law must take into account the time delays due to the finite
propagation speed of the classical fields.

The functional differential equations of motion for classical
field theory are generally difficult, often impossible, to express
in a form that is amenable to analysis. Thus, in order to obtain
useful dynamical predictions from realistic models, it is
frequently to replace the functional differential equations of
motion by approximations that are ordinary or partial differential
equations, see \cite{Ch}. The purpose in these works is to discuss
some of the mathematical issues that must be addressed to obtain a
rigorous foundation for the post--Newtonian dynamics, that is,
Newtonian dynamics with relativistic corrections, see for instance
\cite{Ch} and the references therein. For the electromagnetic
classical field, in the ideal case of a point--charge particle,
the resulting retarded potentials are the Li\'enard--Wiechert
potentials. For the gravitational classical field we must use the
Einstein's field equation. Simple models of these equations are
the subject of current research. The basic idea of post-Newtonian
approximation, from a mathematical point of view, is the expansion
of model equations in powers of $1/c$. From a physical point of
view, the idea is to consider low velocity (compared with the
speed of light). Note, for example, that the relativistic form of
Newton's second law, where the rate of change of the momentum is
given by
\[
\frac{d}{dt} \left( mv(1- \frac{|v|^2}{c^2})^{-1/2} \right) ,
\]
reverts to Newton's law in the low--velocity limit.

According to Maxwell's field equations, a charged particle
produces electromagnetic fields as it moves. Since, in this case,
a particle radiates energy, it must slow down. In the theory of
the electron point charge by considering motion along in a line,
Dirac propose a self-force (the radiation reaction force) given by
\[
F_{\mbox{self}}=\frac{2q^2}{3 c^3} \dddot{x},
\]
which is the half difference of the retarded and advanced forces,
where $q$ is the charge of the electron, see \cite{LL}. Therefore,
a post-Newtonian model for the motion of an electron, confined to
move on a line and with radiation reaction taken into account, is
given by the Abraham--Lorentz equation
\begin{equation}
m \ddot{x}=\frac{2q^2}{3 c^3} \dddot{x}+F, \label{Al}
\end{equation}
where $F$ is an external force. Since the electron radiates
(produces fields that carry energy) the self force should cause
the particle to lose energy and slow down. That's why, the
presence of the third derivative term in the first differential
equation is called radiation damping. However, in these
post-Newtonian models (where the differential equations are not of
second order) the ``runaway" solutions appear, see \cite{LL}. For
instance, in absence of external forces, equation \eqref{Al}
reduces to
\[
m \ddot{x}=\frac{2q^2}{3 c^3} \dddot{x},
\]
and this equation has the solution $\dot{x}=C$ where $C$ is an
arbitrary constant, and other solutions where the acceleration is
proportional to $\exp(3 m c^3t / (2q^2))$. Hence, the acceleration
grows indefinitely with time. This means that, a charge which goes
out of a field, when leaving it, must self-accelerate
indefinitely; which is an absurd. These runaway solutions are
clearly not physical. What do they represent? How should they be
eliminated? and What is the correct Newtonian equation with the
radiation damping taken into account?

The mathematical answer to all these questions is approached in
\cite{Ch} and the subsequent works \cite{Ch2,Ch3}, where these
post-Newtonian models are recognized as singularly perturbed
Newtonian equations. In order to recover the correct Newtonian
equations with the post-Newtonian corrections, the Fenichel's
geometric singular perturbation theory is applied (in particular,
the reduction to the slow--manifold). These Newtonian equations
with the post-Newtonian corrections give physically reasonable
dynamics; in particular, the runaway solutions are eliminated.
Anyway, how can we justify using these models? Note, for instance,
that the slow-manifolds in these models are unstable; nearby
runaway solutions. In applied mathematics, we usually justify
approximations by their stability. To validate the slow--manifolds
reductions it must be shown that the resulting Newtonian model
equations are stable with respect to the dynamics of the original
functional differential equations, the true equations of motion in
classical physics. Therefore further investigations are required
in this direction for the study of the delay equations.

However, it is interesting to note that the presence of a small
delay in a conservative system often results in damped long--term
dynamics on an associated inertial manifold, see \cite{Ch,Ch1}.
For example, the Duffing--type model equation
\[
\ddot{x}+\omega^2 x=-ax(t-\tau)+bx^3(t-\tau),
\]
with small delay $\tau$ in the restoring force, reduce (by a
formal computation to first order in $\tau$) to the van der
Pol--type model equation
\[
\ddot{x}+\tau(3bx^2-a)\dot{x}+(a+\omega^2)x-bx^3=0,
\]
on its inertial manifold. As it has been noticed in \cite{Ch} this
example illustrates a phenomenon that is a reminiscent of
quantization: while most periodic solutions in one parameter
families of periodic solutions in a conservative system disappear
in the presence of a small delay, some persist as limit cycles.
The author of \cite{Ch} asks himself whether this observation has
a physical significance.

The solutions of the functional differential equations can,
however, admit an infinite discrete spectrum. For example, we
consider the retarded harmonic oscillator, given by the linear,
second order, retarded functional differential equation
\[
\ddot{x}+x(t-\tau)=0,
\]
with small delay $\tau$. In \cite{R} it is showed that this
equation exhibits an infinite spectrum of discrete frequencies.
Its general solution is a convergent linear combination of
oscillations at an infinity of discrete (``quantized")
frequencies. As in quantum mechanics, in order to determine a
unique solution an initial function needs to be known. The above
consideration remains valid for any linear functional differential
equation with constant coefficients and constant delay. Moreover,
the locally linear approximation suggests that such
``quantization" is also to be expected for non--linear functional
differential equations.

At the beginning of the 20th century, Planck \cite{PL} initiated
the quantum mechanics with his contribution to the black body
radiation. Einstein \cite{E}, following the ideas of Planck
\cite{PL}, contributed to the development of the theory of quanta
which is the embryonic step needed to arrive to the Quantum
physics theory. It is interesting to note that Poincar\'e had been
implicated in the discussion of the quantum theory, but the
premature death of Poincar\'e deprives of its contributions in
this theory, see for instance \cite{P5,P6}. In fact, Poincar\'e
participated in the first congress of Solvay in October of 1911
and he died in July of 1912. One of the interesting known
phenomenon studied by Poincar\'e is the concept of limit cycle,
see \cite{P1}. This phenomenon does not occur in the Hamiltonian
systems studied by standard physics theories. It only appears in
systems with friction, i.e., systems with dissipative energy.

It is also interesting to note that one of the problems which
originates the quantum theory was the problem that appears when
the idea of the planetarium system is applied to the atomic model.
This idea was proposed by Rutherford in 1911 \cite{RU} as a
consequence of the experimental results obtained when bombing an
atom with $\alpha$ particles. The problem in the model of
Rutherford was that the charged electrons are accelerated in their
movement around the nucleus and by the electromagnetic classical
theory any accelerated charged body radiates energy. The problem
of the atomic stability was initially solved by Bohr in 1913
\cite{Bh} and it marks the success of the quantum theory and its
posterior development. The atomic model of Bohr predicts the
radiation spectrum of certain atoms and the quantization of the
energy in different energy levels is then obtained.

If you see the development of the quantum theory from the initial
contributions, it is evident that each step is made with extra
assumptions. For instance, the introduction of the quanta in the
radiation of a black body and in the foto--electric effect, cf.
\cite{E}, the quantization of the energy in the movement of an
electron which moves as an harmonic oscillator under the influence
of an harmonic restoring force, cf. \cite{PL}. Another example is
the quantization of the angular orbital impulse of an electron in
an atom, although the electron in an atom is accelerated in its
movement around the nucleus. In this last case, it is assumed that
this electron does not radiate energy, see \cite{Bh}. However, we
notice the difference between the Bohr quantization of the angular
orbital impulse of an electron, which moves under the Coulomb
force ($L=n \hbar$ for $n=1,2,3, \ldots$, where $\hbar$ is a
multiple of the Planck constant $h$ divided by $2 \pi$), and the
Planck quantization of the energy of a particle, as an electron,
which moves as an harmonic oscillator ($E=nh\nu$ for $n=1,2,3,
\ldots$). In fact the quantization of the angular orbital impulse
of an electron leads to the quantization of the total energy but
with an equation quite different than the Planck equation.

In \cite{G} it is showed that the intrinsic phenomenon (the
quantization of the energy) that appears in the first and simple
systems initially studied by the quantum theory as the harmonic
oscillator and the movement of a charged particle under the
Coulomb force, can be obtained from the study of dissipative
systems. In other words, it is showed that the phenomenon of the
quantization of the energy of a particle which moves as an
harmonic oscillator can be obtained via a classical system of
equations. The same assertion also applies to the phenomenon of
the quantization of the energy of a charged particle which moves
under the Coulomb force and which loses and wins energy (for
example if we consider the classical case where the electron
radiates and absorbs energy from the electric field of the
nucleus). Therefore, these phenomena are not intrinsic of the
quantum theory, but also appear in classical systems. In fact,
they appear in the qualitative theory of differential equations
developed by Poincar\'e from 1881 \cite{P1}.

Nevertheless, the most important problem is to find the exact form
of the dissipative term and the interpretation of its physical
meaning, see \cite{G}. The retarded case, already explicitly
incorporates certain subtle mathematical features of
electrodynamics and relativity noticed by Poincar\'e, but
overlooked by Einstein and subsequent researchers. Based on the
study of the retarded systems, a simple atomic model given by a
functional differential equation which reproduces the quantized
Bohr atomic model is presented in this paper.

The paper is not at all an alternative to the quantum theory,
because the large development of the quantum theory in all the
past century, the success in all its predictions, is outside of
all doubt. This work does not pretend, in any case, to substitute
quantum mechanics but to complete the knowledge that it gives to
us. On the other hand, the proposed model does not reflect the
whole rich behavior of the quantum modern theories developed from
1925 by Schr\"{o}dinger \cite{SC,SC1,SC2}, Born \cite{B},
Heisenberg \cite{HE}, Dirac \cite{DI}, and others. The goal of the
paper is to ask if the finite propagation speeds is the origin of
the quantum mechanics.

To begin with, in \cite{R}, it is assumed that the two particles
are rotating rigidly in circular orbits around a common center of
masses. Moreover, a force which varies inversely as the square of
the retarded distance is considered. The retarded distance is the
distance from the current position of the electron to the "last
seen" position of the proton. The simple expression for the force
helps us to intuitively understand the consequences of a delay and
under such circumstances, the angular momentum cannot be
conserved. Thus, we have the astonishing situation that, purely
under the action of internal forces, the system suffers a net
torque. Now, the radiation term is introduced. But, the exact form
of the radiation damping term is not clear. Finally, the simple
heuristic case of the retarded inverse square force is used, to
determine whether there can be a balance of forces between the
delay torque and the 3rd order radiation damping. A total success
does not take place and a value of $r$ which is smaller than the
Bohr radius is obtained. Nevertheless, the author, in \cite{R},
affirms that further investigations are required to determine the
exact effects of radiative damping, and that it was prematurely
concluded that radiative damping makes the classical hydrogen atom
unstable. In the following section we present a simple atomic
model based on a functional differential equation which reproduces
the quantized Bohr atomic model. It is important to stand out that
what we will carry out in the following section is not a
post--Newtonian approach in which $\tau$ is small. This is what
has been made up to now and in the mentioned works
\cite{Ch,Ch1,Ch2,Ch3}. From now on, we will accept that the laws
governing the movement have a delay (a delay that does not need to
be small) and we will find a solution of the functional
differential equation in a very simple case.

\section{The retarded electrodynamic 2-body problem}

\begin{figure}[htb]
\centerline{\hbox{
\epsfig{file=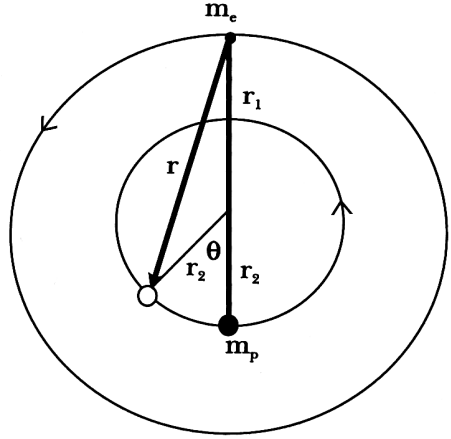,height=6.0cm,width=6.0cm} }} \caption{The
retarded electrodynamic 2-body problem.} \label{fig1}
\end{figure}
\par

We consider two particles interacting through the retarded inverse
square force. The force on the electron exerted by the proton is
given by
\begin{equation}
{\bf F}=K \frac{e^2}{r^3} \ {\bf r}. \label{fo}
\end{equation}
The force acts in the direction of the 3--vector ${\bf r}$, along
which the proton is "last seen" by the electron. The 3--vector
${\bf r}$ may be represented by
\[
{\bf r}= {\bf r}_p(t-\tau) - {\bf r}_e(t),
\]
where ${\bf r}_p(t)$ and ${\bf r}_e(t)$ denote respectively the
instantaneous position vectors of the proton and electron,
respectively, at time $t$, and $\tau$ is the delay, so that ${\bf
r}_p(t-\tau)$ is the "last seen" position of the proton. Assuming
that the two particles are in rigid rotation with constant angular
velocity $\omega$, and referring back to Fig. 1, we have, in
3--vector notation,
\[
{\bf r}_e=r_1[\cos \omega t \ {\bf \hat{\i}} + \sin \omega t \
{\bf \hat{\j}}],
\]
and
\[
{\bf r}_p=-r_2[\cos \omega (t-\tau) \ {\bf \hat{\i}} + \sin \omega
(t-\tau) \ {\bf \hat{\j}}].
\]
Hence, the 3--vector ${\bf r}$ is given by
\[
{\bf r}=[-r_2\cos \omega (t-\tau) -r_1\cos \omega t] \ {\bf
\hat{\i}} + [-r_2 \sin \omega (t-\tau) -r_1 \sin \omega t] \ {\bf
\hat{\j}},
\]
Now, we introduce the polar coordinates $(r, \theta)$ and define
the unitary vectors ${\bf l}= \cos \theta \ \hat{\i} + \sin \theta
\ \hat{\j}$ and  ${\bf n}= - \sin \theta \ \hat{\i} + \cos \theta
\ \hat{\j}$. By straightforward calculations it is easy to see
that the components of the force \eqref{fo} in the polar
coordinates are
\[
F_r=K \frac{e^2}{r^3}{\bf r} \cdot {\bf l} = (-r_2 \cos (\omega
\tau) - r_1) K \frac{e^2}{r^3} \,
\]
and
\[
F_\theta=K \frac{e^2}{r^3}{\bf r} \cdot {\bf n} = r_2 \sin (\omega
\tau) K \frac{e^2}{r^3} \,
\]
The equations of the movement are
\begin{eqnarray}
m \ddot{r}- m r \dot{\theta}^2 = F_r, \label{F1} \\
m r \ddot{\theta} + 2 m \dot{r} \dot{\theta} = F_\theta.
\label{F2}
\end{eqnarray}
The second equation \eqref{F2} can be written in the form
\begin{equation}
\frac{1}{r}\frac{d L}{dt}= \frac{1}{r}\frac{d}{dt}(m r^2 \dot
\theta) = F_\theta=  r_2 \sin (\omega \tau) K \frac{e^2}{r^3}.
\label{FF2}
\end{equation}

In 1913 Bohr \cite{Bh} introduced the quantization of the angular
momentum of the form $L=n h/ (2 \pi)$ where $h$ is the Planck
constant. If we accurately study equation \eqref{FF2} we see that
the analytic function $\sin (\omega \tau)$ has a numerable number
of zeros given by
\begin{equation}
\omega \tau = k \pi \, , \label{eeqq}
\end{equation}
with $k \in \Z$, which are stationary orbits of the system of
equations \eqref{F1} and \eqref{F2}. When $\omega \tau \ne k \pi$
we have a torque which conduces the electron to the stationary
orbits without torque, that is, with $\omega \tau =k \pi$.

This is a new form of treating the hydrogen atom from a dynamic
point of view instead of from a static point of view, as it has
been made up to now. Moreover, in this model the delay $\tau$ is
not small, in fact
\[
\tau = \frac{k \pi}{\omega}= \frac{k \pi}{2 \pi/ T} = \frac{k
T}{2} \, ,
\]
where $T$ is the time taken by an electron to complete its orbit.
Therefore, the delay is a multiple of the half--period $T/2$.

On the other hand, in a first approximation, the delay $\tau$ can
be equal to $r/c$ (the time that the field uses to goes from the
proton to the electron at the speed of the light). In this case,
from equation \eqref{eeqq} we have
\begin{equation}
\tau= \frac{k \pi}{\omega} = \frac{r}{c} \, . \label{eeqq2}
\end{equation}
Taking into account that $\omega = v_{\theta} /r$, from
\eqref{eeqq2} we have $v_{\theta}/ c = k \pi$. However, from the
Relativity theory we know that $v_{\theta}/ c < 1$, then we must
introduce a new constant $g$ in the delay. Hence, $\tau= g \, r/c$
and the new equation \eqref{eeqq2} is
\begin{equation}
\tau= \frac{k \pi}{\omega} = \frac{g \, r}{c} \, , \label{eeqq3}
\end{equation}
and now $v_{\theta}/ c = k \pi / g$, i.e. $v_{\theta}= k \pi c /
g$ and from \eqref{eeqq3} we also have $r= k \pi c /(g \omega)$.
In our model case of a classical rigid rotation we have $\theta=
\omega t$ with $\omega
> 0$. Therefore, $\dot \theta = \omega$ and $\ddot{\theta}=0$.
Hence, equation \eqref{F2} for $\omega \tau = k \pi$ is
\[
2 m \dot{r} \omega =0,
\]
which implies $\dot{r}=0$ and $r=r_k$ where $r_k$ is a constant
for each $k$. On the other hand, equation \eqref{F1} for $\omega
\tau = k \pi$ takes the form:
\begin{equation}
-m r \dot \theta^2 = -m \frac{v_{\theta}^2}{r} = (-r_2(-1)^n-r_1)
K \frac{e^2}{r^3} \approx -r K \frac{e^2}{r^3}, \label{F3}
\end{equation}
assuming that $r \sim r_1$ due to $r_2 \ll r_1$.

From the definition of angular momentum $L=m r^2 \dot \theta= m
r^2 \omega = m r v_{\theta}$ we have that $v_{\theta}=L/(m r)$.
Substituting this value of $v_{\theta}$ into equation \eqref{F3}
we obtain $r=L^2/(m K e^2)$. The energy of the electron
(substituting the values of $v_{\theta}$ and $r$) is given by
\begin{equation}
E=\frac{mv_{\theta}^2}{2}- K \frac{e^2}{r}=-\frac{K^2 m \, e^4}{2
\, L^2}. \label{F5}
\end{equation}
The angular momentum for $\omega \tau = k\pi$ is
\begin{equation}
L= m v_{\theta} r = m \frac{k \pi c}{g} \frac{L^2}{m K e^2},
\label{F4}
\end{equation}
which is an equation for the angular momentum. Isolating the value
of $L$ we obtain $L=(K e^2 g)/(k \pi c)$. If we introduce this
value of the angular momentum in the expression of the energy
\eqref{F5} we have
\[
E= -\frac{m \pi^2 c^2}{2 g^2} k^2,
\]
In 1890 Johannes Robert Rydberg generalized Balmer's formula and
showed that it had a wider applicability. He wrote his formula as
\[
\frac{1}{\lambda} = R \left( \frac{1}{n_1^2}-\frac{1}{n_2^2}
\right)
\]
where $\lambda$ is the wavelength, $n_1$ and $n_2$ are integer
numbers and $R$ is known as the Rydberg constant. Therefore, the
energy levels are proportional to $1/n^2$ and, of course,
negative, because these are bound states, and we count energy zero
from where the two particles are infinitely far apart. Hence,
(identifying $n=|k|$) we must impose that the constant $g=k^2g_1$
where $g_1$ is another new constant and then the energy takes the
form
\[
E= -\frac{m \pi^2 c^2}{2 g_1^2} \frac{1}{k^2},
\]
If we recall the expression of the energy levels given by Bohr in
1913
\[
E=-\frac{me^4}{8 \varepsilon_0^2h^2n^2},
\]
we can compare these two expressions of the energy levels and we
get the explicit value of the Planck constant
\begin{equation}
h=\frac{e^2 g_1}{2 \pi \, \varepsilon_0 \, c}. \label{F6}
\end{equation}
The dimensional analysis gives $[J \cdot s]= [C^2]/([C^2/(J
m)][m/s])$ which is correct. The introduction of a new fundamental
constant is avoided (as it happens in the quantum mechanics with
the Planck constant) because through the delay the speed of the
light $c$ appears. The appearance of this dimensional constant
$c$, usually absent in the non--relativistic quantum mechanics,
allows to give the expressions of the physical quantities with the
correct dimensions. Substituting into equation \eqref{F6} the
value of the electron charge $e$, the electric permittivity
constant $\varepsilon_0$ and the value of the speed of light $c$
we obtain that the adimensional constant $g_1$ must take the value
$g_1=429.868$.

We notice that this value of $g_1=429.868$ is the inverse value of
the fine structure constant $ 1/\alpha=137.036$ multiplied by
$\pi$, that is, $g_1=\pi/\alpha$. Moreover, the value of this fine
structure constant $\alpha$ is
\[
\alpha= \frac{e^2}{4 \pi \epsilon \hbar c},
\]
and when substituting this value of $g_1$ in the explicit
expression of the Planck constant \eqref{F6} the equality is
identically satisfied. Therefore we have found the value of the
adimensional constant $g_1$ and consequently the expression of the
delay $\tau$ which is
\begin{equation}
\tau= \frac{g \, r}{c}= k^2 \, \frac{g_1 r}{c}= k^2 \, \frac{\pi
r}{\alpha c}. \label{re}
\end{equation}

From the found value of the angular momentum and the value of
$v_{\theta}= \alpha c /k$ we have
\begin{equation}
L= \frac{K e^2 g}{k \pi c}=\frac{K e^2 k}{\alpha c} = m v_{\theta}
r = m \frac{\alpha c}{k} r. \label{F7}
\end{equation}
Isolating the value of $r$ from equation \eqref{F7} we obtain
\[
r=\frac{K e^2 k^2}{m \alpha^2 c^2}.
\]
Taking into account that $K=1/(4 \pi \varepsilon_0)$ and the value
of the fine structure constant $\alpha$, we arrive to the
classical radii of the stationary orbits
\begin{equation}
r_n=\frac{h^2 \varepsilon_0 \, n^2}{\pi \, m \, e^2}. \label{rr}
\end{equation}
Moreover, the relation $v_{\theta}/c= k \pi /g= \alpha/ k$ is
consequent with the definition of $\alpha$. One of the
interpretations of the fine structure constant $\alpha$ is that
$\alpha$ relates the speed of the electron in the lowest energy
level in the atom of hydrogen with the speed of the light. This is
straightforward because if we substitute the expression \eqref{rr}
of $r_n$ in the classical expression of $v_{\theta}=n h /(2 \pi m
r_n)$ given by the quantum mechanics, for the case $n=1$, and we
divide by the speed of the light $c$, we obtain that
$v_{\theta}/c= \alpha$.

Summarizing, we could have begun our analysis with the found delay
definition \eqref{re}, because our model reproduces the Bohr atom
faithfully. Quantum mechanics in the Bohr atom is in fact the
first approximation in the value $v/c$ of the delay.

The fact that the Planck constant is expressed in function of the
parameters associated to the particular model system (see
\eqref{F6}) could be a problem. If we consider another model
system, usually another expression of the Planck constant would
appear and another numerical value for the constant $g$ would have
to be chosen, and hence another expression for the delay. However,
we think that in the expression of the delay, in fact, we will
obtain the same constant $g$ in all the problems  as it happened
in the development of the quantum mechanics (perhaps not exactly
the same constant but a constant that will be expressible in terms
of known constants and in function of the fine structure
constant). Suppose, for a moment, that we are in 1913 and we made
this work in that date. In order to be consistent with the
Relativity theory, a new constant $g$ needs to appear. This
constant $g$ seems to be another important constant of the nature.
Suppose that we follow this line of research and we apply this
theory to other systems. What we think that it would have happened
is that in the resolution of these new systems, the constant $g$
would also appear again and all the physicians would have thought
that this $g$ is a new important constant of the nature (in fact
it is $k^2 \pi / \alpha$). If we do not compare with the Bohr
model, we find at the end that $v/c = \pi/(g_1 n )$ and $\pi/g_1$
would be the relation between the velocity of the electron at the
first energy level and the speed of the light (this relation is
known in quantum mechanics as the fine structure constant
$\alpha$).

In the following section we will obtain a first approximation of
the fine structure corrections of Sommerfeld \cite{So,So2} to the
hydrogen atom.

\section{The fine structure of the hydrogen atom}

In old quantum mechanics, one of the more spectacular successes
which helped to accept the Sommerfeld-Wilson-Ishiwara quantization
was the study realized by Sommerfeld in \cite{So,So2}. These works
treat with the hydrogenoid atoms giving an explanation of the fine
structure of the hydrogen atom discovered by Michelson
\cite{M1,M2}. Sommerfeld applied Special Relativity theory
assuming that the electron inside the atom is travelling near the
speed of light. He had obtained the following correction to the
energy levels
\begin{equation}
E \simeq - \frac{1}{2} \, \mu \, (Z \alpha c)^2 \frac{1}{n^2}
\left [ 1+ \frac{\alpha^2 Z^{\,2}}{n} \left (
\frac{1}{n_\psi}-\frac{3}{4n} \right) \right] \ , \label{F8}
\end{equation}
where $n$ is a positive integer number and $n_\psi=1,2,3, \ldots,
n.$ The agreement of this formula with the experimental data is
fortuitous, because the correct formula is obtained using the
Dirac equation, where the spin of the electron is taken into
account. The final expression is the same that \eqref{F8} changing
$n_\psi$ by $j+1/2$ where $j$ is the total angular momentum of the
electron and $j$ can take the following values for the $n$ level
$j=1/2, 3/2, \ldots, n-1/2$.

In fact, the fine structure is a result of relativistic
corrections to the Schr\"{o}dinger equation, derived from the
relativistic Dirac equation for an electron of mass $m$ and charge
$e$ in an external electrical field $ -\nabla \Phi({\bf r})$.
Performing an expansion in $v/c$, the result for the Hamiltonian
$\hat{H}$ can be written as $\hat{H}=\hat{H}_0+ \hat{H}_1$, where
\[
\hat{H}_0=- \frac{\hbar^2}{2m}\Delta - \frac{Ze^2}{4\pi
\varepsilon_0 r}
\]
is the non-relativistic hydrogen atom, $Z=1$, and $\hat{H}_1$ is
treated as a perturbation to $\hat{H}_0$, using perturbation
theory. $\hat{H}_1$ consists of three terms: the kinetic energy
correction, the Darwin term, and the spin-orbit coupling,
$\displaystyle \hat{H}_1 = \hat{H}_{\rm KE} + \hat{H}_{\rm Darwin}
+ \hat{H}_{\rm SO}.$

Using the delay theory introduced in this work we are going to
obtain a first approach to the fine structure of the hydrogen
atom, as straightforward application of the result of the previous
section.

The relativistic kinetic energy is given by $E_{K}= \mu c^2 -
\mu_0 c^2$, where $\mu_0$ is the rest mass and $\mu=\mu_0/
\sqrt{1-v^2/c^2}$. Hence, the total energy is given by
\[
E= \mu_0 c^2 \left[ \frac{1}{\sqrt{1-v^2/c^2}} -1 \right] - K
\frac{e^2}{r},
\]
or (substituting the value of the Potential energy in terms of
angular momentum)
\[
E= \mu_0 c^2 \left[ \frac{1}{\sqrt{1-v^2/c^2}} -1 \right] -
\frac{K^2 \mu e^4}{L^2}.
\]
From the value of the angular momentum $L=(K e^2 k^2)/(k \alpha
c)$ of the previous section, we obtain
\[
E= \mu_0 c^2 \left[ \frac{1}{\sqrt{1-v^2/c^2}} -1 \right] -
\frac{\mu \alpha^2 c^2}{k^2} \, .
\]
Substituting $\mu=\mu_0/ \sqrt{1-v^2/c^2}$ and developing in
powers of $v/c$ to fourth order we get
\[
E \simeq  \frac{1}{2} \mu_0 c^2 \left[ \frac{v^2}{c^2} +
\frac{3}{4} \frac{v^4}{c^4} + \ldots \right]- \frac{\mu_0 \alpha^2
c^2}{k^2} \left[ 1 + \frac{v^2}{2c^2} + \frac{3}{8}
\frac{v^4}{c^4} + \ldots \right] .
\]
Finally, substituting the value of $v= \alpha c /k$ (obtained in
the previous section) we arrive to the expression
\[
E \simeq - \frac{1}{2} \, \mu \, (\alpha c)^2 \frac{1}{k^2} \left
[ 1+ \frac{\alpha^2 }{k^2} \left ( 1-\frac{3}{4} \right) \right],
\]
and, identifying $n=|k|$, it takes the form
\[
E \simeq - \frac{1}{2} \, \mu \, (\alpha c)^2 \frac{1}{n^2} \left
[ 1+ \frac{\alpha^2 }{n} \left ( \frac{1}{n}-\frac{3}{4n} \right)
\right].
\]
If we compare this last expression with \eqref{F8} we see that we
have obtained only a first approach to the Sommerfeld quantization
because we only obtain the case $n_\psi=n$. In this first
approximation we have not introduced the intrinsic angular
momentum of the electron (spin). Therefore further investigations
are required to find the correct model of the delay equations with
the intrinsic angular momentum of the electron taken into account.

\section{Concluding remarks}

The aim of my work is the understanding of which can be the
explanation of the quantic behavior of nature. Physicians, working
in quantum mechanics, believe that this explanation lies on the
same theory. However, there is no impediment to the existence of a
deeper description giving as an observable result the effects of
quantum mechanics. In order to give an example, this is the same
thing that happens with classical mechanics and statistical
mechanics of $n$ bodies. Under the laws of statistical mechanics,
the laws of classical mechanics lie, that govern the movement of
each one of the $n$ bodies. This work does not pretend, in any
case, to substitute quantum mechanics but to complete the
knowledge that it gives to us.

The whole foundation of quantum mechanics, starting from
Heisenberg and Schr\"{o}dinger is somewhat ad hoc. This is not a
great deficiency, since any theory must start with some
assumptions, and the less is the number of such assumptions, the
better is a theory. In this way, the model that we present in this
work is better than the Bohr model because the unique assumption
is that the electrodynamic interaction has finite propagation
speed. The introduction of some model, different from the existing
one, should go accompanied with a possibility of describing the
new phenomena, yet to be discovered. We hope to give an answer to
new phenomena in the context of this theory in future works.

From the fact that the atomic Bohr model can be completely
described by means of functional differential equations, we
believe that in the future, observing the physical reality at a
deeper level, we will be able to interpret the laws of probability
of the quantum physics as the statistical results of values of
certain variables perfectly determined that at the moment are
hidden for us...(we know that this idea has been extremely
discussed) the history of science shows us that the current state
of the knowledge is always provisional and that it should exist,
beyond what is known, immense regions to be discovered. However,
functional differential equations are fundamentally different from
ordinary differential equations, and their solutions can have
qualitative features which are impossible for solutions of
ordinary differential equations. The physical consequences of
these differences are explained at length in \cite{R2}. We mean
that the rich features that can appear in functional differential
equations may explain quantum mechanics, from a deeper point of
view.\\

\noindent{\bf Acknowledgements:}

The author would like to thank Prof. H. Giacomini from
Universit\'e de Tours and Prof. M. Grau from Universitat de Lleida
for several useful conversations and remarks.


\begin{thebibliography}{99}


\bibitem{Bh}{\sc N. Bohr}, {\it On the Constitution of Atoms and
Molecules}, Phil. Mag. {\bf 26} (1913), 1--25.

\bibitem{B} {\sc M. Born}, {\it Zur Quantenmechanik der
Sto\ss vorg\"ange}, Z. Phys., {\bf 37} (1926), 863--867.

\bibitem {Ch} {\sc C. Chicone}, {\it What are the equations of
motion of classical physics?}, Can. Appl. Math. Q. {\bf 10}
(2002), no. 1, 15--32.

\bibitem {Ch1} {\sc C. Chicone, S.M. Kopeikin, B. Mashhoon and D. Retzloff}, {\it Delay
equations and radiation damping}, Phys. Letters {\bf A 285}
(2000), 17--16.

\bibitem {Ch2} {\sc C. Chicone}, {\it Inertial and slow manifolds for
delay equations with small delays}, J. Differential Equations {\bf
190} (2003), no. 2, 364--406.

\bibitem {Ch3} {\sc C. Chicone}, {\it Inertial flows, slow flows, and
combinatorial identities for delay equations}, J. Dynam.
Differential Equations {\bf 16} (2004), no. 3, 805--831.

\bibitem{DI} {\sc P.A.M. Dirac}, {\it The quantum theory of the
electron}, Proc. Roy. Soc. {\bf A 117} (1928), 610--624.

\bibitem{E} {\sc A. Einstein}, {\it \"Uber einen die Erzeugung und Verwandlung
des Lichts betreffenden heuristischen Gesichtspunkt}, Ann. Phys.
{\bf 17} (1905), 132--148.

\bibitem{G} {\sc J. Gin\'e}, {\it \ On the classical descriptions of the
quantum phenomena in the harmonic oscillator and in a charged
particle under the coulomb force}, Chaos Solitons Fractals {\bf
26} (2005), 1259--1266.

\bibitem{HE} {\sc W. Heisenberg}, {\it \"Uber den anschaulichen Inhalt der
quantentheoretischen Kinematik und Mechanik}, Z. Phys., {\bf 43}
(1927), 172--198.

\bibitem{LL} {\sc L.D. Landau and E.M. Lifshitz}, {\it The
classical theory of fields}, Oxford: Pergamon Press, 1971.

\bibitem{M1} {\sc A.A. Michelson}, {\it On the application of interference
methods to spectroscopic measurements I}, Phil. Mag. {\bf 31}
(1892), 338--346.

\bibitem{M2} {\sc A.A. Michelson}, {\it On the application of interference
methods to spectroscopic measurements II}, Phil. Mag. {\bf 34}
(1892), 280--299.

\bibitem{PL} {\sc M. Planck}, {\it \"Uber das Gesetz der Energieverteilung im
Normalspectrum}, Ann. Phys. {\bf 4} (1901), 553-563.

\bibitem{P1} {\sc H. Poincar\'e}, {\it M\'emoire sur les courbes
d\'efinies par les \'equations diff\'erentielles.} Journal de
Math\'ematiques {\bf 37} (1881), 375-422; {\bf 8} (1882), 251-296;
Oeuvres de Henri Poincar\'e, vol. I, Gauthier-Villars, Paris,
(1951), pp. 3-84.

\bibitem{P5} {\sc H. Poincar\'e}, {\it Sur la th\'eorie des
quanta}, C. R. Acad. Sci. Paris {\bf 153} (1911), 1103--1108.

\bibitem{P6} {\sc H. Poincar\'e}, {\it Sur la th\'eorie des quanta},
J. Physique Th\'eorique et Appliqu\'ee, 5 s\'erie {\bf 2} (1912),
5--34.

\bibitem{R} {\sc C.K. Raju}, {\it The electrodymamic 2-body problem
and the origin of quantum mechanics}, Foundations of Physics {\bf
34} (2004), 937--962.

\bibitem{R2} {\sc C.K. Raju}, {\it Time: towards a consistent theory},
Kluwer academic, Dordrecht, 1994.

\bibitem{RU} {\sc E. Rutherford}, {\it The Scattering of $\alpha$ and
$\beta$ Particles by Matter and the Structure of the Atom}, Phil.
Mag. {\bf 21} (1911) 669--668.

\bibitem{SC} {\sc E. Schr\"{o}dinger}, {\it Quantisierung als Eigenwertproblem. (Erste
Mitteilung.)}, Ann. Phys. (Leipzig) {\bf 79} (1926), 361--376.

\bibitem{SC1} {\sc E. Schr\"{o}dinger}, {\it Quantisierung als Eigenwertproblem.
(Zweite Mitteilung.)}, Ann. Phys. (Leipzig) {\bf 79} (1926),
489--527.

\bibitem{SC2} {\sc E. Schr\"{o}dinger}, {\it Quantisierung als Eigenwertproblem.
(Dritte Mitteilung.)}, Ann. Phys. (Leipzig) {\bf 80} (1926),
437--490.

\bibitem{So} {\sc A. Sommerfeld}, {\it Einf\"urung in die Quantentheorie, Oscillator und
Rotator}, Atombau und Spectrallinien, Chapter 2 \S 3, Friedr.
Vieweg \& Sohn, Braunschweig, 1924.

\bibitem{So2} {\sc A. Sommerfeld}, {\it Die Bohrsche Theorie der Balmerserie}, Atombau und
Spectrallinien, Chapter 2 \S 4, Friedr. Vieweg \& Sohn,
Braunschweig, 1924.

\end{thebibliography}
\end{document}